\documentclass[a4paper,11pt]{article}
\usepackage{pos}

\usepackage{slashed}
\usepackage{color}
\usepackage{multicol}
\usepackage{wrapfig}
\usepackage{graphicx}  % needed for figures
\usepackage{dcolumn}   % needed for some tables
\usepackage{cancel}
\usepackage{graphicx}
\usepackage{wrapfig}
\usepackage{subfig}
\usepackage{float}
\usepackage{tikz}
\usepackage[all]{xy}
\usepackage{wrapfig}
\usepackage{bbold}

\title{Studying chiral imbalance using Chiral Perturbation Theory.}
\author*[a]{Andrea Vioque-Rodríguez}
\author[a]{Angel Gómez Nicola}
\author[b]{Domenec Espriu}

\affiliation[a]{Departamento de F\'{\i}sica Te\'orica and IPARCOS. Univ. Complutense.}

\affiliation[b]{Departament of Quantum Physics and Astrophysics and Institut de Ci\'encies del Cosmos (ICCUB), Universitat de Barcelona.}

\emailAdd{avioque@ucm.es}
\emailAdd{gomez@fis.ucm.es}
\emailAdd{espriu@icc.ub.edu}

\abstract{We analize the most general low-energy effective lagrangian including local parity violating terms parametrized by an axial chemical potential $\mu_5$. This result is obtained following the external source method, up to $\mathcal{O}(p^4)$ order in the chiral expansion for two light flavours. We show that the $\mathcal{O}(p^4)$ lagrangian includes new terms proportional to $\mu_5^2$ and new low-energy constants. Finally, the $\mu_5$ and temperature dependences of several observables related to the vacuum energy density are studied.}

\FullConference{%
  *** Particles and Nuclei International Conference - PANIC2021 ***\\
  *** 5 - 10 September, 2021 ***\\
  *** Online ***
}

\begin{document}
\maketitle

\section{Introduction}

A convenient way to parametrize a Parity-breaking source or chiral imbalance is by means of a constant axial chemical potential $\mu_5$ to be added to the QCD action over a given finite space-time region. The chiral charge may remain approximately conserved during the fireball evolution in a typical heavy-ion collision, giving rise in the light quark sector to a chemical potential term which is equivalent to consider an axial source $a_{\mu}^0=\mu_5\delta_{\mu 0}$ in the QCD generating functional.

We provide in \cite{Espriu:2020dge} a model-independent approach, constructing the most general effective lagrangian for the lightest degrees of freedom in the presence of the $\mu_5$ source. Preliminary ideas along this line have
been proposed in \cite{Andrianov:2019fwz}. In this contribution we will review the main phenomenological consequences in terms of observables such as the energy density, the pion decay constant, the pion mass, the quark condensate and the topological susceptibilities.

\section{Effective lagrangian}

The construction of the most general, model-independent, effective lagrangian can be carried out within the framework of the external source method \cite{Gasser:1984gg}. To do that, the so called "spurion" fields $Q_{R,L}(x)$ are introduced, where in our case $Q_L=-Q_R=(\mu_5/F)\mathbb{1}$ and $F$ is the pion decay constant in the chiral limit. There are additional terms in the effective lagrangian depending on $Q_{R,L}$. Moreover, the operator $\text{tr}(U^{\dagger}d_{\mu}U)$ has to be considered, unlike in standard ChPT where that operator vanishes. To $\mathcal{O}(p^2)$, the only modification to the chiral lagrangian is a constant term:
\begin{equation}
	\mathcal{L}_2=\dfrac{F^2}{4}\text{tr}\left[\partial_{\mu}U^{\dagger}\partial^{\mu}U+2B_0\mathcal{M}\left(U+ U^{\dagger}\right)\right]+2\mu_5^2F^2\left(1-Z+\kappa_0\right).
\end{equation}
with $Z$ defined in \cite{Knecht:1997jw}.

At $\mathcal{O}(p^4)$ we have new terms constructed out of the $Q$ operators and $\text{tr}(U^{\dagger}d_\mu U)$. The possible contributions (including $Q$ operators) are of the form: $ddQQ, \chi QQ, QQQQ$, and the explicit $\mu_5$ corrections are:
\begin{equation}
		\mathcal{L}_4(\mu_5)=\mathcal{L}_4^0(\mu_5=0)+\kappa_1\mu_5^2\text{tr}\left(\partial_\mu U^{\dagger}\partial^\mu U\right)+\kappa_2\mu_5^2\text{tr}\left(\partial_0 U^{\dagger}\partial^0 U\right)+\kappa_3\mu_5^2\text{tr}\left(\chi^\dagger U+\chi U^{\dagger}\right)+\kappa_4\mu_5^2.
\end{equation}
where these $\kappa_i$ constants above can be compared with the electromagnetic LEC $k_i$ given in \cite{Knecht:1997jw} by taking from the general lagrangian $Q_R=Q_L=Q$ with $Q$ the electromagnetic charge matrix.

\section{Physical consequences}

\begin{figure}[H]
	\begin{center}\includegraphics[width=6cm]{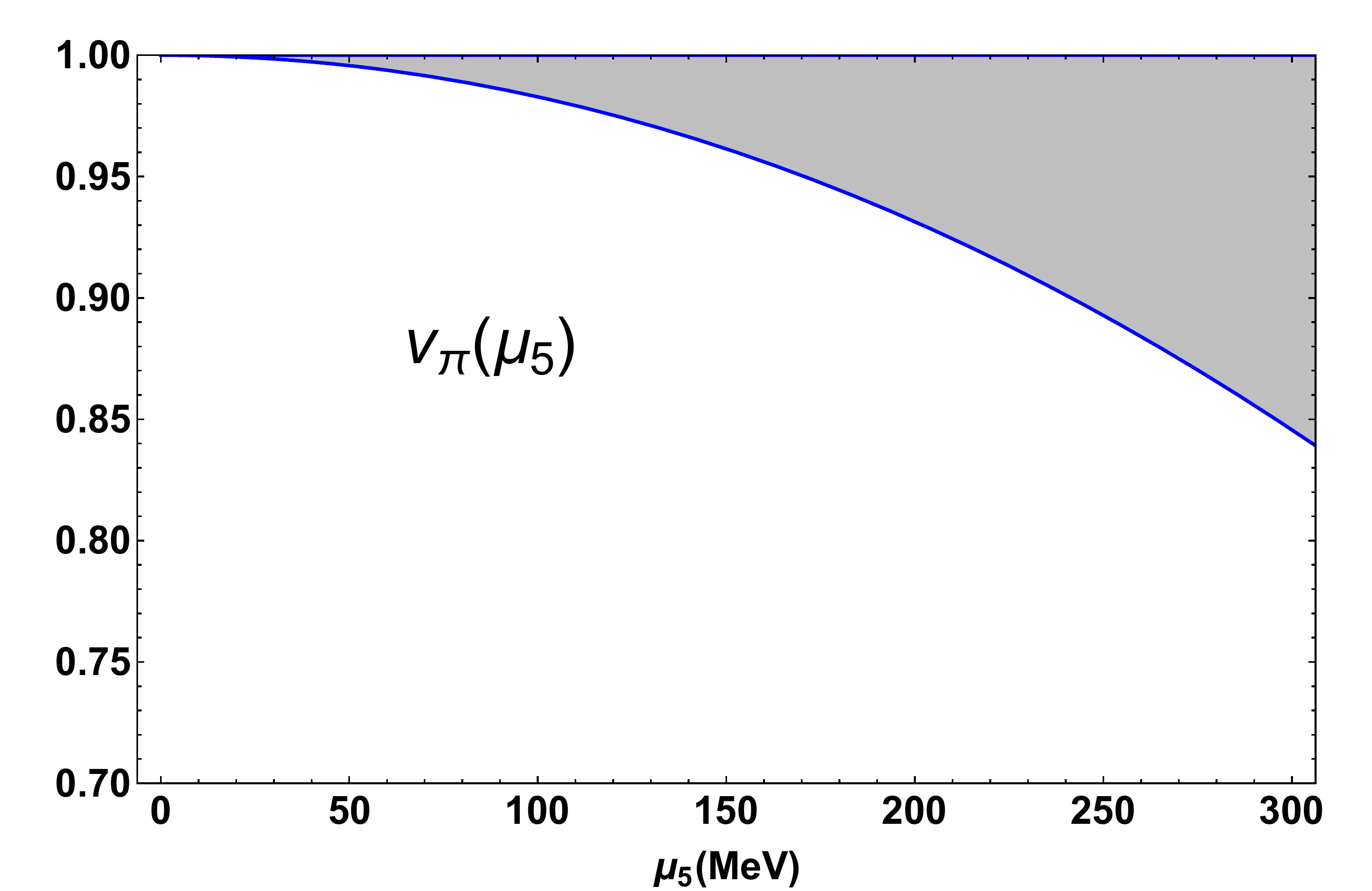}
	\includegraphics[width=6cm]{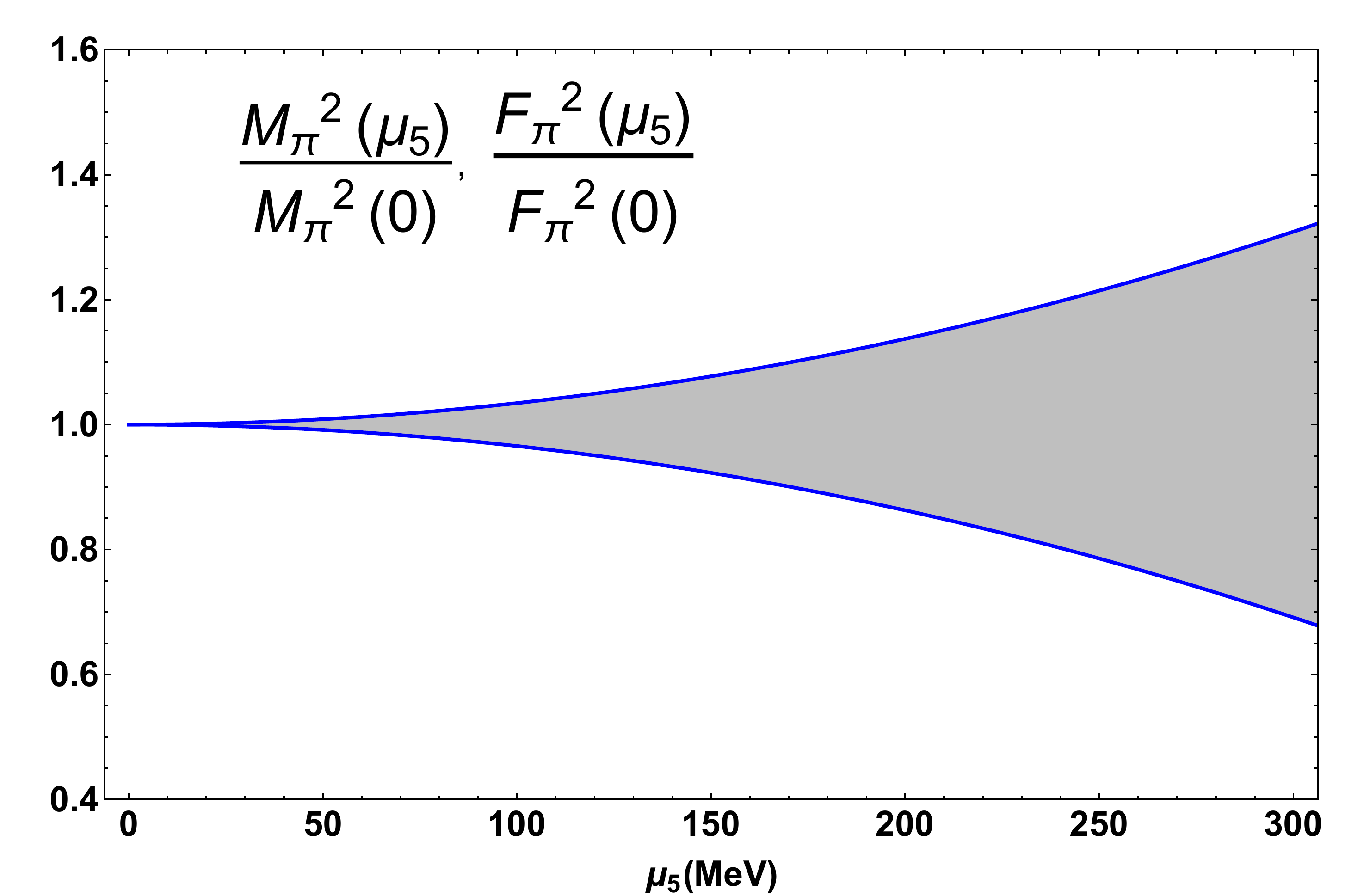}
	\vspace{-0.2cm}
	\caption{$\mu_5$  dependence of pion velocity, pion mass and pion decay constant, to leading order in ChPT. The grey bands correspond to the uncertainties of the LEC within natural values $1/(16\pi^2)$ which is their expected size from loop corrections.}
	\label{massfpi}
	\end{center}
\end{figure}

The external field breaks manifest Loretz invariance and the spatial and time components of the pion decay constant are different \cite{Pisarski:1996mt}. The two main physical consequences of that are the pion velocity and pion mass. Requiring that the pion velocity remains smaller than
the speed of light for any $\mu_5$, the constrain $\kappa_2<0$ is obtained. On the other hand, in order to the square pion mass remains positive, $\kappa_1-\kappa_3<0$. However, a decreasing pion mass for low values of $\mu_5$ does not necessarily imply a tachyonic mode so this last condition may be too restrictive. We plot in Figure \ref{massfpi} the dependence of $v_\pi$ and $M_\pi^2$ with $\mu_5$ expected within natural values for the LECs.
\vspace{-0.3cm}

At order $\mathcal{O}(p^2)$ in the vacuum energy density only contributes the constant part of the $\mathcal{L}_2$ lagrangian. The $\mathcal{O}(p^4)$ includes the contribution from the kinetic part of $\mathcal{L}_2$ and the field-independent contributions coming from $\mathcal{L}_4$. Finally, to $\mathcal{O}(p^6)$ the contributions are the two-loop closed diagram with four-pion vertices coming from $\mathcal{L}_2$, the $\mathcal{O}(\pi^2)$ part of $\mathcal{L}_4$ and the field-independent contributions coming from $\mathcal{L}_6$.

\begin{figure}[H]
		\begin{center}
			\includegraphics[width=6.4cm]{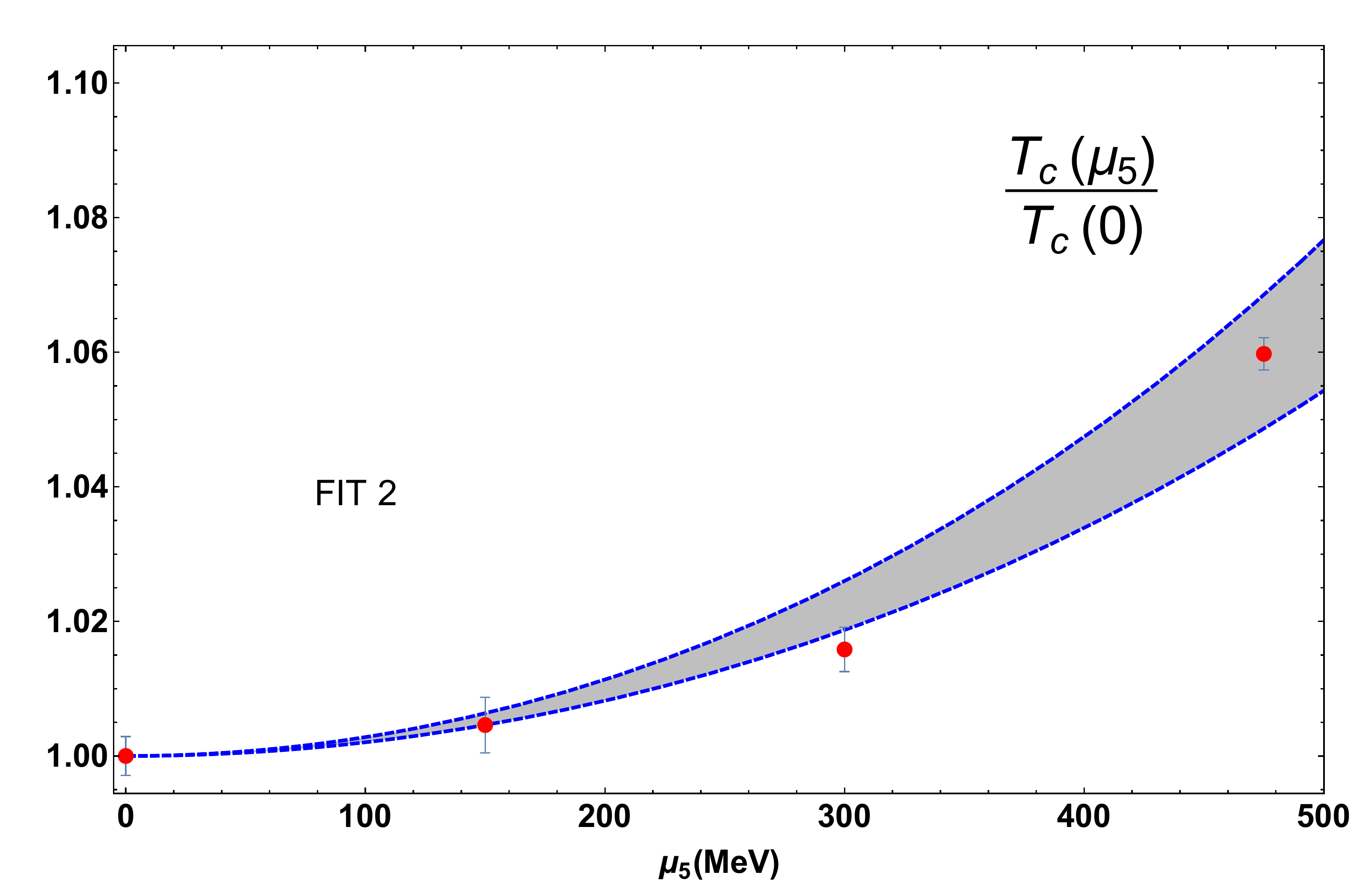}
			\includegraphics[width=6.4cm]{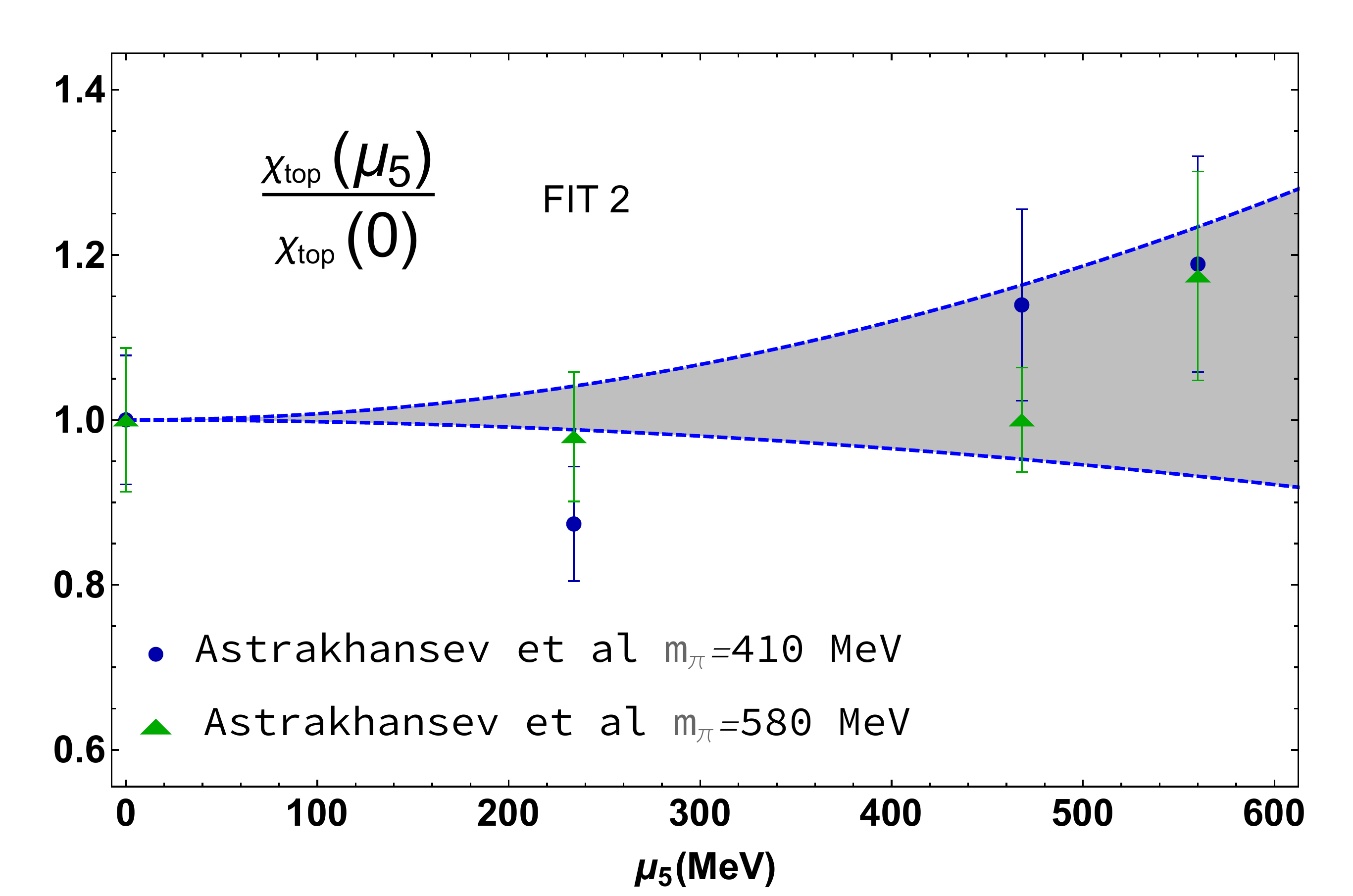}
		\end{center}
	    \vspace{-0.5cm}
		\caption{Left: Fit of the critical temperature, $T_c(\mu_5)/T_c(0)$, for three lattice points in the chiral limit. We have used the three points with lower $\mu_5$ ( and $\mu_5\neq 0$) obtained in \cite{Braguta:2015zta}. Right: Fit of lattice data to $\chi_{top}(\mu_5)/\chi_{top}(0)$ where the number of lower $\mu_5$ points with $\mu_5\neq 0$ considered is 3 ($m_\pi=410$ MeV)+3 ($m_\pi=580$ MeV).}
		\label{tcritmu5chitop}
\end{figure}

The main features of chiral symmetry restoration can be read from the quark condensate and the scalar susceptibility which are derived from the vacuum energy density. The first $\mu_5$ correction to the quark condensate is temperature independent and its behavior is determined by the sign of $\kappa_3$, it increases if $\kappa_3>0$ and it decreases  if $\kappa_3<0$. Lattice results clearly show growing condensate and $T_c$ with $\mu_5$ \cite{Braguta:2015owi}. For the ratio $\langle \bar{q}q\rangle_l^{\text{NNLO}}(T,\mu_5)/\langle \bar{q}q\rangle_l^{\text{NNLO}}(0,\mu_5)$ the $\kappa_i$ dependence reduces to the combinations $\kappa_a=2\kappa_1-\kappa_2$ and $\kappa_b=\kappa_1+\kappa_2-\kappa_3$ (although in the chiral limit it depends only on $\kappa_a$). In order to provide more
quantitative conclusions we plot the $\mu_5$ dependence of the critical temperature normalized by its value at $\mu_5=0$, this dependence is just quadratic in $\mu_5$ and is given in \cite{Espriu:2020dge}. This curve for the physical pion mass lies very close to the chiral limit one and the lattice points \cite{Braguta:2015zta} clearly fall into the uncertainty given by the natural values range of $\kappa_a$ and $\kappa_b$. We compared a fit with two and three points (left side of Figure \ref{tcritmu5chitop}) in the chiral limit with a fit in the massive case fixing the $\kappa_a$ parameter. These fits show that the chiral limit approach with just one parameter $\kappa_a$ is a robust approximation.

The dependence of the topological susceptibility, $\chi_{top}$, with low and moderate $\mu_5$ is controlled by the $\kappa_3$ constant. The fit of that observable to the lattice data, which we can see in the right side of Figure \ref{tcritmu5chitop}, shows that the results for $\kappa_3$ are compatible with zero and the error bands are much narrower than the natural values for this constant.  The lattice points used for the fit are those in \cite{Astrakhantsev:2019wnp}.

Another quantity studied in the lattice is the chiral charge density, $\rho_5(\mu_5)$. We perform a fit of $\rho_5(\mu_5)$ to the lowest values of $\mu_5$ provided in \cite{Astrakhantsev:2019wnp}. The simple linear dependence fits very well the lowest $\mu_5$ lattice points. The prediction for $\kappa_0$ is consistent with the fit allowing the other parameters to be free. An interesting result is that the analysis of the energy density does not favor a $\mu_5\neq 0$ minimum for the free energy for moderate values of $\mu_5$.

Finally, the thermodynamic pressure ($P$) and the speed of sound ($c_s$) have also been calculated. The $\mu_5$ corrections to $P$ are parametrized by $\kappa_2$ and $\kappa_b$ but in the chiral limit, which corresponds to the ultrarelativistic free pion gas, only the $\kappa_2$ term survives. On the other hand, $c_s$ depends only on $\kappa_b$ and as can be seen in \cite{Espriu:2020dge}, when we include the $\mu_5$ corrections, $c_s^2$ remains below $1/3$. The uncertainty band for $\kappa_b$ actually narrows as $T$ increases.

\section{Conclusions}

In this work we have analyzed the effective chiral lagrangian for nonzero chiral imbalance up to fourth order and from that we have study several observables and the dependence of their $\mu_5$ corrections with the $\kappa_i$ constants. Besides this we have compared our results with existing lattice data.

\textit{Acknowledgements} Work partially supported by research contracts FPA2016-75654-C2-2-P, FPA2016-76005-C2-1-P, MDM-2014-0309
(Ministerio de Econom\'ia y Competitividad), 2017SGR929 (Generalitat de Catalunya) and the European Union Horizon 2020 research. A. V-R acknowledges support from a fellowship of the UCM predoctoral program.

\end{document}